\documentclass[fleqn,usenatbib,useAMS]{mnras}
\usepackage{float}
\usepackage[T1]{fontenc}
\usepackage{ae,aecompl}
\usepackage{graphicx}	
\usepackage{epstopdf}
\usepackage{amssymb}
\usepackage{supertabular}
\usepackage{siunitx}
\usepackage{caption}
\usepackage{amsmath}	
\usepackage{multicol}        
\usepackage{bm}		
\usepackage{pdflscape}	
\usepackage{geometry}
\usepackage{multirow}

\def\beq{\begin{eqnarray}}
\def\eeq{\end{eqnarray}}

\title{Reincarnations of massive stars in active galactic nucleus discs}

\author[J.-T. Xing et al.]{Jing-Tong Xing$^{1}$, Tong Liu$^{1}$\thanks{Contact e-mail: \href{mailto:tongliu@xmu.edu.cn}{tongliu@xmu.edu.cn}}, Jiao-Zhen She$^{2}$%
\\
$^{1}$Department of Astronomy, Xiamen University, Xiamen, Fujian 361005, China
\\
$^{2}$School of Electrical Engineering, Tongling University, Tongling 244000, China}

\pubyear{2026}

\begin{document}
\label{firstpage}
\pagerange{\pageref{firstpage}-\pageref{lastpage}}
\maketitle

\begin{abstract}
The origin and evolution of massive stars in active galactic nucleus (AGN) discs remain uncertain. We develop a semi-analytical model that follows the evolution of an embedded core-collapse supernova (CCSN) remnant and the subsequent formation and growth of a compact gas cloud. In the dense disc environment, efficient radiative cooling can strongly compress or bypass the Sedov-Taylor stage and drive the remnant rapidly into radiative snowplow evolution. After the remnant loses its interior pressure support, partial backflow of cooled shell fragments and refilling gas may initialize a pressure-confined and tidally limited seed cloud. The cloud then grows through shear-limited Hill capture until the gravitational, tidal, shear, photoionization, and magnetic conditions for collapse are simultaneously satisfied. The outcome depends strongly on the supermassive black-hole (SMBH) mass and explosion radius. Models with the lowest SMBH mass yield fewer than one massive star per event on average, whereas the most massive SMBH models can produce from several to several hundred. For a top-heavy initial mass function, massive stars dominate the resulting stellar mass and feedback budget. Embedded supernovae may therefore provide a localized gas-recycling channel for second-generation massive-star formation in AGN discs.
\end{abstract}


\section{Introduction} \label{sec: intro}

The gravitational energy released by supermassive black holes (SMBHs) through accretion of surrounding matter is regarded as the standard model of the central engine of active galactic nuclei \cite[AGNs; e.g.][]{2012agn..book.....B}. Although there are still many problems, this theory has been widely accepted and supported by multi-band observational evidence. The AGN unified model's core structure consists of three main parts \citep{1984ARA&A..22..471R}, the SMBH with a mass range of  ${10}^6-{10}^{10}M_\odot$, and the gaseous accretion disc surrounding the SMBH, as well as the dust ring that can block part of the radiation and produce anisotropic observation characteristics. The extreme physical environment of high temperature, high pressure, and high density inside the accretion disc makes it a natural laboratory for studying high-energy astrophysical processes, including instantaneous burst phenomena, such as tidal disruption events \cite[TDEs; e.g.][]{1999ApJ...514..180U}, gamma-ray bursts \cite[GRBs; e.g.][]{1999ApJ...521..502C}, star formation (SF) and evolution under extreme conditions \citep[e.g.][]{2025ApJ...981...16F}, gravitational wave (GW) sources \cite[such as binary star mergers; e.g.][]{2011ApJ...726...28B}, and accretion processes of compact objects \citep[e.g.][]{2024ApJ...966L...9Z,2025ApJ...991..167X}.

\begin{figure*}
    \centering
    \includegraphics[width=0.7\linewidth]{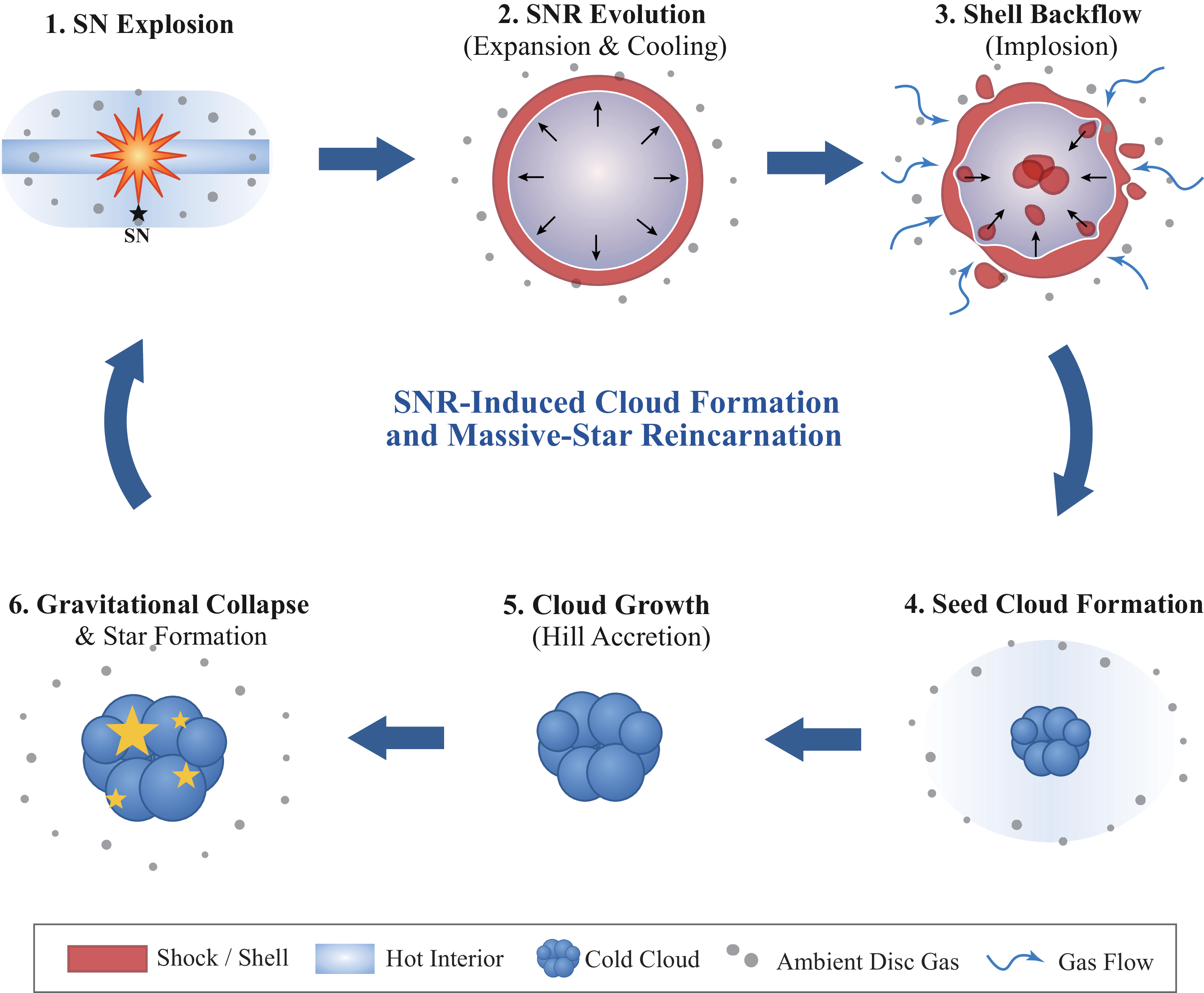}
    \caption{Schematic illustration of the proposed SNR-induced gas-recycling and star-formation channel in an AGN disc. The remnant evolves through the hydrodynamic and radiative expansion phases before the loss of interior pressure support allows partial backflow of cooled shell material and refilling disc gas. A compact seed cloud may subsequently form, grow through local Hill accretion, and collapse into second-generation stars. The backflow is schematic and is not intended to represent the coherent contraction of a complete spherical shell.}
    \label{fig1}
\end{figure*}

Reverberation mapping (RM) observations provide important constraints on the geometry and kinematics of the broad-line region (BLR) \citep{1993PASP..105..247P}, while spectroscopic studies indicate that quasar BLRs can be metal-rich \citep{2022ApJ...925..121W}. This suggests that there may be an efficient element enrichment mechanism within the disc, and this mechanism may be closely related to SF and stellar evolution as well as various burst phenomena in the dense environment. Due to the AGN disc's high-density and highly turbulent environment, the SF and supernovae (SNe) processes within it are significantly different from those in the interstellar medium \cite[ISM;][]{2023ApJ...950..161L}. Theoretical research and numerical simulation results predict that the strong turbulent environment within the disc will inhibit the formation of low-mass stars \citep{1999A&A...344..433C,2008Sci...321.1060B,2008A&A...477..419C}, resulting in the internal initial mass function (IMF) of stars possibly being of the ``top-heavy" type \citep{2009MNRAS.394..191H,2022MNRAS.512.2573T}, that is, the proportion of high-mass SF will be higher than that of low-mass SF \citep{2022MNRAS.517.6205P,2023MNRAS.519..397O}. Meanwhile, the heavy metal elements mainly produced by SNe Ia, core-collapse SNe (CCSNe), and black hole (BH) hyperaccretion can also reasonably explain the strong metallic abundance of BLR \citep{2010ApJ...719L.148W,2021ApJ...920....5L,2022ApJ...934....1Q}.

For the stars and compact objects embedded in the AGN accretion disc, they will migrate inward or outward due to the density perturbation of the disc gas, and the migration pattern and the ratio of the mass of the migrating material to the mass of the SMBH are related \citep{1991MNRAS.250..505S,2010MNRAS.401.1950P,2016ApJ...819L..17B}. Therefore, the dense environment of AGN discs is expected to yield a higher CCSN rate than a dilute environment. High star density and a gas-rich climate can also increase the formation efficiency of binary star systems \citep{1999ApJ...521..502C}. These binary stars may eventually merge \citep{2011ApJ...726...28B,2021ApJ...914L..19Z} and become one of the GW sources within the AGN disc \citep{2012MNRAS.425..460M}.

More and more evidence indicates that massive stars formed by disc gravitational instability exist in the AGN BLR \citep{1980SvAL....6..357K,1999A&A...344..433C,2008A&A...477..419C}. They evolve rapidly within the disc, forming accretion-modified stars (AMS) and long-lived even immortal stars \citep{2023ApJ...954...84W,2025ApJ...981...16F}, and eventually generate CCSNe, thereby increasing the high metallicity at the centre of the galactic nuclei. However, the formation mechanism of these massive stars remains unclear \citep{1999A&A...344..433C,2007ARA&A..45..565M}, and the evolutionary processes of SN remnants (SNRs) after SNe are also challenging to elucidate. The analysis and numerical simulation results show that the SNR evolution timescale within the AGN disc is shorter compared to that in the ISM environment \citep{1980ApJ...237..769S,1992MNRAS.255..713T,2021ApJ...906...15M}. The three-dimensional (3D) simulation by \cite{2021ApJ...906...15M} shows that after the SNR undergoes a century of evolution, there will be gas falling back into its central cavity. \cite{2024ApJ...965..168R} demonstrated that the gas reflux of SNR can trigger implosion, form massive clouds, and eventually trigger gravitational instability to form stars \citep{2024ApJ...971L..44R}. This mechanism may also exist in the AGN disc and become one of the origins of massive stars. However, research on this process in AGN discs remains scarce and urgently needs to be studied.

\begin{figure*}
    \centering
    \includegraphics[width=0.9\linewidth]{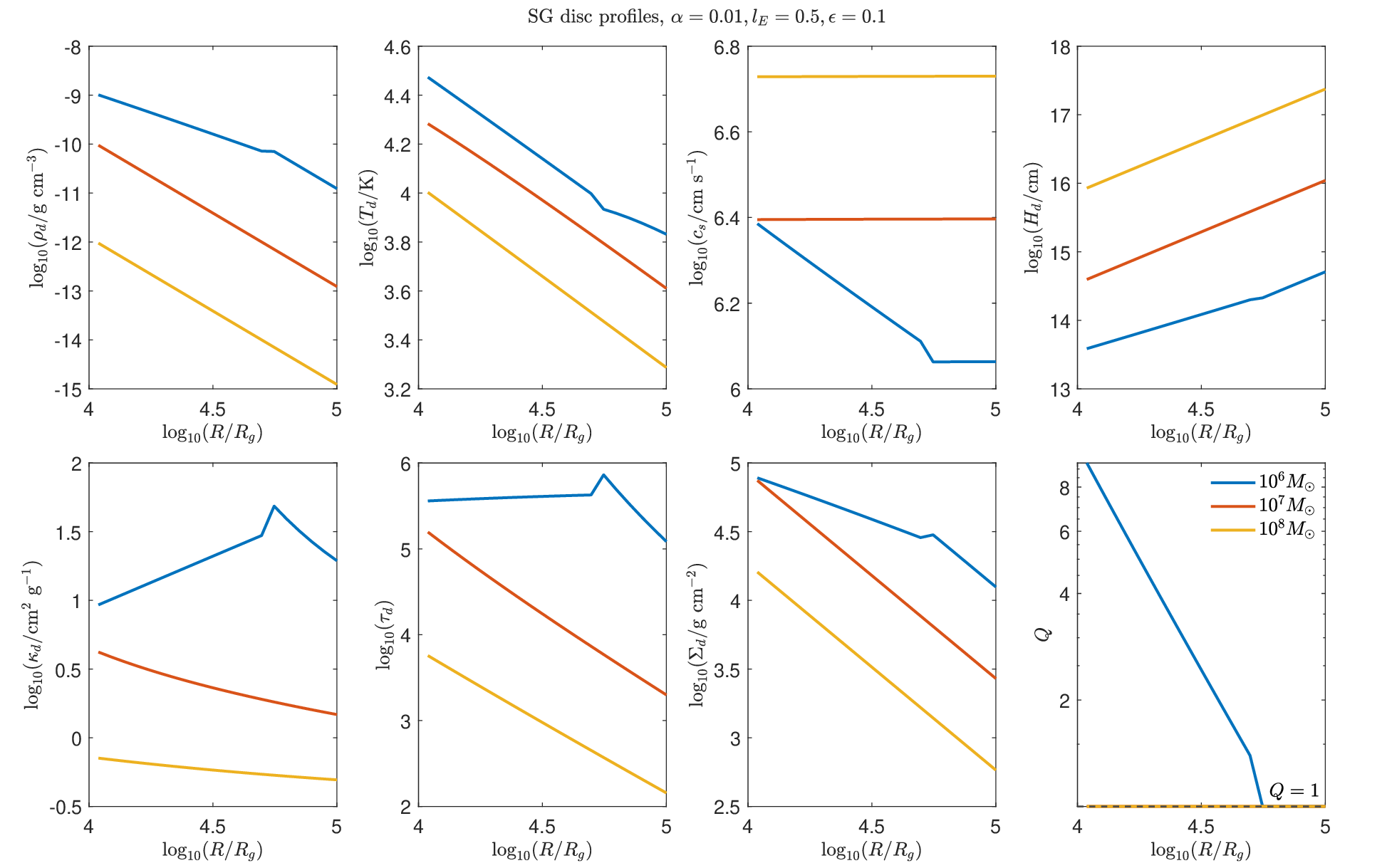}
    \caption{Profiles of AGN disc quantities: density $\rho_\mathrm{d}$, temperature $T_\mathrm{d}$, sound speed $c_\mathrm{s}$, scale height $H_\mathrm{d}$, opacity $\kappa_\mathrm{d}$, optical depth $\tau_\mathrm{d}$, surface density $\Sigma_\mathrm{d}$, and Toomre parameter $Q$, shown from left to right and from top to bottom. Blue, orange, and yellow curves correspond to $M_\bullet=10^6,10^7$, and $10^8M_\odot$, respectively. The dashed horizontal line in the $Q$ panel marks $Q=1$.}
    \label{fig2}
\end{figure*}

In this paper, we develop a semi-analytical model for SNR-triggered SF in AGN discs, as summarized schematically in Figure \ref{fig1}. Efficient radiative cooling drives the remnant into the snowplow phases. After the outward shock stalls and the hot interior loses pressure support, the ambient disc pressure may drive partial backflow of cooled shell fragments and refilling gas into the cavity. A small fraction of this material initializes a compact seed cloud, which subsequently grows through local Hill accretion and may eventually collapse. This channel is intended as a localized gas-recycling mechanism rather than a dominant global mode of SF.

The structure of this paper is as follows. In Section~\ref{sec: Initial condition}, we introduce the adopted AGN-disc model, the SN and progenitor parameters, and the local ambient conditions used for the SNR calculation. In Section~\ref{sec: star formation}, we follow the evolution of the SNR from its early expansion to radiative shell formation, implosion, and seed-cloud formation. We then model the subsequent growth and gravitational collapse of the seed cloud and estimate the resulting population of second-generation massive stars under a top-heavy IMF. The main conclusions and remaining uncertainties are summarized in Section~\ref{sec: Summary}.

\section{Model setup and initial conditions}\label{sec: Initial condition}

We use the AGN disc model proposed by \cite{2003MNRAS.341..501S} (SG). This model addresses the issue of gravitational instability ($Q<1$) of the standard accretion disc by \cite{1973A&A....24..337S} at large radii $R\gtrsim2\times10^3 R_\mathrm{g}$ by invoking additional heating in the disc, where $R_\mathrm{g}=2GM_\mathrm{\bullet}/c^2$ is the Schwarzschild radius, $G$ is the gravitational constant, $M_\mathrm{\bullet}$ is the SMBH mass, $c$ is the speed of light, and $Q$ is the Toomre parameter by \cite{1964ApJ...139.1217T}. Since the subsequent analysis of SNR mainly focused on $R > 10^4 R_\mathrm{g}$, we only plotted the evolution of the disc parameters in this region. Our subsequent analysis relies on the numerical results depicted in Figure \ref{fig2}. The viscosity parameter is $\alpha=0.01$, the Eddington ratio $l_\mathrm{E}=0.5$, and the radiative efficiency $\epsilon=0.1$. In the following calculations, we consider three SMBH masses, $M_\bullet=10^6,10^7$, and $10^8M_\odot$, where $M_\odot$ represents the Sun's mass. The detailed description of the disc model can be found in \cite{2003MNRAS.341..501S}.

\begin{table*}
\centering
\begin{tabular}{lcccccc}
\hline
Stage & $t$ [yr] & $R_\mathrm{in}/R_\mathrm{g}$ & $R_\mathrm{out}/R_\mathrm{g}$ 
& $\rho_\mathrm{out}/\rho_\mathrm{in}$ 
& $P_\mathrm{gas,out}/P_\mathrm{gas,in}$ 
& $T_\mathrm{out}/T_\mathrm{in}$ \\
\hline
FREE & $3.21\times10^{-2}$ & 29926.0 & 30074.0 & 0.985 & 0.982 & 0.996 \\
ST   & $4.98\times10^{-2}$ & 29911.5 & 30088.5 & 0.982 & 0.978 & 0.996 \\
PDS  & 3.75                 & 29695.9 & 30304.1 & 0.941 & 0.927 & 0.985 \\
MCS  & 6.84                 & 29646.6 & 30353.4 & 0.932 & 0.916 & 0.983 \\
\hline
\end{tabular}
\caption{Radial environmental asymmetry across the SNR shell for the fiducial case with $M_\bullet=10^8~M_\odot$ and $R_\mathrm{I}=10^{4.5}~R_g$. Here $R_\mathrm{in}$ and $R_\mathrm{out}$ denote the shell positions on the sides closer to and farther from the SMBH, respectively. Here $P_\mathrm{gas}=\rho_\mathrm{d}k_\mathrm{B}T_\mathrm{d}/(\mu_0m_\mathrm{p})$.}
\label{tab1}
\end{table*}

Considering a massive star initially located at $R_\mathrm{I}$, it may come from the gravitational capture by the disc or be directly formed under local gravitational instability of the outer disc. Similar to planetary migration in the protoplanetary disc \citep{1996Icar..124...62P,2010apf..book.....A}, the stars migrate due to density perturbations in the disc gas \citep{2016ApJ...819L..17B}. \cite{2023ApJ...950..161L} proposes that the migration velocity $v_\mathrm{m}$ can be simplified to
\begin{equation}
    v_\mathrm{m}=-\frac{2}{M_{\star}} f \Sigma_\mathrm{d} \Omega R_\mathrm{I}^{3}\left(\frac{R_\mathrm{I}}{H_\mathrm{d}}\right)^{2}\left(\frac{M_{\star}}{M_{\mathrm{\bullet}}}\right)^{2},
    \label{eq1}
\end{equation}
where $\Sigma_\mathrm{d}$ is the surface density of the AGN disc, $\Omega=(GM_\mathrm{\bullet}/R_{\mathrm{I} }^3)^{1/2}$ is the angular velocity of the star's orbit, $f\approx0.02$ is the numerical factor \citep{2023ApJ...950..161L}, $M_{\star}$ is the star mass, and $H_\mathrm{d}$ is the half scale height of the disc. We define the local migration timescale as $t_\mathrm{mig}=R_\mathrm{I}/|v_\mathrm{m}|$, which is listed in Table~\ref{tab2} as a consistency check against the SNR evolution time.

The SG disc indicates that for the outer region of the disc where $Q<1$, the disc will rapidly fragment locally to form clumps \citep{2007MNRAS.374..515L}. The fragmentation timescale is very short, typically close to the orbital dynamical time, which is much shorter than the AGN lifetime \citep{2023ApJ...954...84W}. During an AGN episode, another plausible channel is that pre-existing stars from the nuclear star cluster become embedded via star-disc interactions/capture and subsequently undergo accretion-driven growth in the dense disc environment \citep{2020MNRAS.499.2608F,2021ApJ...916L..17W}. The probability of stars being captured is very low \citep{2024MNRAS.528.4958W}, and in environments with high shear and large relative velocities (inner disc or high-temperature regions), the efficiency of star accretion and merger will be greatly reduced, resulting in a possible long timescale for the growth of massive stars \citep{2025ApJ...987..188C}. A plausible and likely efficient channel for producing massive stars in AGN discs is the direct gravitational collapse of disc clumps, although stellar capture, mergers, and accretion-driven growth may also contribute in some regions of parameter space. These massive stars are expected to end their evolution as CCSNe within the AGN lifetime.

We consider a red supergiant (RSG) $\sim 25~M_\mathrm{\odot}$, with a radius of $\sim500~R_\mathrm{\odot}$, and an effective temperature of $4,000~\rm{K}$, where $R_\odot$ represents the Sun's radius. Three representative values for the initial location of the star $R_\mathrm{I}$, i.e., $10^{4}~R_\mathrm{g}$, $10^{4.5}~R_\mathrm{g}$, and $10^{5}~R_\mathrm{g}$, are adopted to ensure that the star generates SNe within the migration timescale. When a star undergoes an SN explosion at its initial position $R_\mathrm{I}$, the SNR shell samples a finite radial range of the AGN disc as it expands. We use a cylindrical coordinate system and consider only the radial coordinate in the disc midplane. The radial position of a point on the SNR shell is written as $R(\theta,t)=R_\mathrm{I}+R_\mathrm{m}(t)+R_\mathrm{SNR}(t)\sin\theta$, where $R_\mathrm{m}(t)\simeq v_\mathrm{m}t$ is the radial displacement caused by stellar migration, $R_\mathrm{SNR}(t)$ is the forward-shock radius, and $\theta\in[-\pi/2,\pi/2]$ parameterizes the radial projection of the shell. The two limiting values, $\theta=-\pi/2$ and $\theta=\pi/2$, correspond to the inner and outer sides of the remnant, i.e., the sides closer to and farther from the SMBH, respectively. This construction is used only to estimate the radial variation of the ambient disc properties across the remnant, and is not intended to describe the full 3D geometry of the SNR.

To quantify the possible radial environmental asymmetry, we compare the disc properties at the inner and outer edges of the remnant for different evolutionary stages. The result is summarized in Table~\ref{tab1}. For the fiducial case with $R_\mathrm{I}=10^{4.5}R_\mathrm{g}$, the radial span of the remnant remains small compared with $R_\mathrm{I}$. Even at the momentum-conserving snowplow (MCS) stage, the full radial extent is only about $2.4\%$ of the central radius. The corresponding contrasts between the outer and inner sides are also modest: the density, gas pressure, and temperature ratios are $\rho_\mathrm{out}/\rho_\mathrm{in}\simeq0.93$, $P_\mathrm{gas,out}/P_\mathrm{gas,in}\simeq0.92$, and $T_\mathrm{out}/T_\mathrm{in}\simeq0.98$, respectively. Therefore, the local disc environment sampled by the two radial sides of the SNR is only weakly asymmetric. Such a mild radial gradient is not expected to qualitatively change the stage-by-stage evolution of the SNR. In the following analysis, we therefore adopt the disc properties evaluated at the explosion radius $R_\mathrm{I}$ as the fiducial ambient conditions: $\rho_\mathrm{d}\equiv \rho_\mathrm{d}(R_\mathrm{I}), T_\mathrm{d}\equiv T_\mathrm{d}(R_\mathrm{I}),n_\mathrm{d}\equiv {\rho_\mathrm{d}}/{\mu_0m_\mathrm{p}}$. The resulting model should therefore be regarded as a local, representative SNR evolution in the AGN disc, rather than as two independent inner-side and outer-side calculations.

\section{Cloud formation and reincarnations of massive stars } 
\label{sec: star formation}

\subsection{Evolution of SNRs}
\label{subsec: The evolution of the SN remnant}

\begin{table*}
\centering
\small
\renewcommand{\arraystretch}{1.25}
\begin{tabular}{lccccccccc}
\hline
$R_\mathrm{I}$ ($R_\mathrm{g}$)
& $t_\mathrm{FREE}$ (yr)
& $t_\mathrm{ST}$ (yr)
& $t_\mathrm{PDS}$ (yr)
& $t_\mathrm{MCS}$ (yr)
& $t_\mathrm{mig}$ (yr)
& $R_\mathrm{MCS}$ (AU)
& $H_\mathrm{d}$ (AU)
& $R_\mathrm{MCS}/H_\mathrm{d}$
& $P_\mathrm{orb}$ (yr) \\
\hline
$10^{4}$ 
& 0.0135 & 0.0097 & 0.782 & 1.37 & $3.50\times10^8$
& 218 
& 498 
& 0.438 
& 277 \\

$10^{4.5}$ 
& 0.0426 & 0.0636 & 5.06 & 8.94 & $3.51\times10^9$
& 919 
& $2.81\times10^{3}$ 
& 0.327 
& $1.56\times10^{3}$ \\

$10^{5}$ 
& 0.135 & 0.416 & 32.8 & 58.3 & $3.51\times10^{10}$
& $3.88\times10^{3}$ 
& $1.58\times10^{4}$ 
& 0.245 
& $8.77\times10^{3}$ \\
\hline
\end{tabular}
\caption{Characteristic SNR scales for the fiducial $M_\bullet=10^8M_\odot$ models, where $R_\mathrm{MCS}$ and $H_\mathrm{d}$ are in Astronomical Units (AU). Comparing $t_\mathrm{MCS}$ with $P_\mathrm{orb}$ shows that the early SNR evolution is local compared with orbital shear. The ratios $R_\mathrm{MCS}/H_\mathrm{d}<1$ indicate that the remnant remains within the local disc scale height before the end of the strong SNR expansion phase. At $R_\mathrm{I}=10^4R_\mathrm{g}$, $t_\mathrm{ST}<t_\mathrm{FREE}$, indicating that the nominal ST stage is strongly compressed.}
\label{tab2}
\end{table*}

To make the hierarchy of relevant scales explicit, we list the characteristic SNR time and length-scales in Table \ref{tab2}. The SNR evolution time is always much shorter than the local orbital period. For the fiducial $M_\bullet=10^8M_\odot$ cases, the small value of $t_\mathrm{MCS}/P_\mathrm{orb}$ ensures that the expanding shock and the subsequent shell backflow operate on timescales orders of magnitude shorter than the local dynamical shearing timescale. Consequently, the rapid, supersonic hydrodynamic evolution of the SNR remains unaffected by the galactic tidal shear during its active expansion phase. The same table also shows that the innermost models have $t_\mathrm{ST}<t_\mathrm{FREE}$, so the Sedov-Taylor (ST) phase is strongly compressed or effectively skipped rather than forming a long adiabatic stage.

Similar to SNR evolution in the ISM \citep{2015A&A...576A..95I,2015ApJ...802...99K}, the time evolution of the SNR radius follows distinct power-law scalings that characterize four main stages of the expansion \citep{1988ApJ...334..252C,2011piim.book.....D,2015ApJ...802...99K}. Free expansion stage (explosion radius $\propto t$), ST stage ($\propto t^{2/5}$), pressure-driven snowplow (PDS) stage ($\propto t^{2/7}$), and MCS stage ($\propto t^{1/4}$). The SNR evolution ends when the shock velocity becomes comparable to the velocity dispersion of the ambient medium and the shock merges with the surrounding environment \citep{1988ApJ...334..252C,2011piim.book.....D,2013MNRAS.433.1970F,2018MNRAS.477.2716K}. We set the explosion energy as $E_\mathrm{SN} = 10^{51} ~\rm{erg}$ and the ejecta mass as $M_\mathrm{ej} \approx 2 ~M_\mathrm{\odot}$. The adopted ejecta mass represents a fallback-dominated core-collapse event after pre-SN mass loss, rather than the full initial stellar mass. The evolutionary cut-off times of the four stages are calculated as follows.

\textit{Free expansion stage}: We assume that the SN ejecta are macroscopically isotropic and that the early remnant can be approximated as a locally spherical blast wave. The free-expansion stage ends when the swept-up disc mass becomes comparable to the ejecta mass,
\begin{equation}
    \frac{4\pi}{3}\rho_\mathrm{d}R_\mathrm{FREE}^3 = M_\mathrm{ej}.
    \label{eq2}
\end{equation}
The corresponding transition time is
\begin{equation}
    t_\mathrm{FREE}
    =\left(\frac{3M_\mathrm{ej}}{4\pi\rho_\mathrm{d}}\right)^{1/3}
    \frac{1}{v_\mathrm{ej}} .
    \label{e3}
\end{equation}

We adopt the commonly used broken power-law density structure of SN ejecta \citep{1982ApJ...258..790C,2023ApJ...950..161L}. Based on the principles of mass and energy conservation, the velocity of the ejecta's break surface is $v_\mathrm{break} = \sqrt{2(5-m)(n-5)E_\mathrm{SN} / [(3-m)(n-3)M_\mathrm{ej}]}$. Assuming homologous expansion, the maximum velocity of the outermost ejecta is scaled by the dimensionless break radius $x_0 \sim 0.3$ \citep{2023ApJ...950..161L}, yielding:
\begin{equation}
    v_\mathrm{ej}
    =
    \frac{1}{x_0}
    \left[
    \frac{2(5-m)(n-5)E_\mathrm{SN}}
    {(3-m)(n-3)M_\mathrm{ej}}
    \right]^{1/2},
    \label{eq4}
\end{equation}
where $m = 0$ and $n = 7$ respectively represent the density distribution of the inner region with approximately uniform density, and the outer region that drops sharply \citep{2023ApJ...950..161L}.

\textit{ST stage}: After the reverse shock heats the ejecta, the evolution of the SNR for $t>t_\mathrm{FREE}$ is well described by the evolution of an idealized, point source explosion because the pressure of the SNR far exceeds the disc pressure. The forward-shock radius of SNR could be solved as $R_\mathrm{ST}=\xi(E_\mathrm{SN}t^2/\rho_\mathrm{d})^{1/5}$ \citep{2011piim.book.....D}. We thus have the immediate post-shock temperature during the ST stage given by
\begin{equation}
    T_\mathrm{ST}
    =
    \frac{3}{16}
    \frac{\mu_0 m_\mathrm{p}}{k_\mathrm{B}}
    v_\mathrm{ST}^2
    =
    \frac{3\xi^2}{100}
    \frac{\mu_0 m_\mathrm{p}}{k_\mathrm{B}}
    \left(
    \frac{E_\mathrm{SN}}{\rho_\mathrm{d}}
    \right)^{2/5}
    t^{-6/5},
    \label{eq5}
\end{equation}
where $\xi\sim1.35$ is the constant of SNR radius \citep{1959sdmm.book.....S,1967pswh.book.....Z}, and $k_\mathrm{B}$ is the Boltzmann constant. We assume that the chemical composition of the disc gas is dominated by hydrogen and helium, and when calculating the shock wave post-shock temperature in the ST stage, we adopt the fully ionized approximation. We take the mean molecular weight to be $\mu_0 = 0.618$ \citep{2017ApJ...846..133K,2025ApJ...987..188C}, so that the mean particle mass is $\mu_0 m_\mathrm{p}$. This approximation applies to the high-temperature stage ($T \ge 10^6~\rm{K}$) after the shock, and the UV/X-ray radiation background of AGN will further maintain a high degree of ionization. The cooling of SNRs embedded in AGN discs proceeds through distinct regimes. At very early times, particularly in the inner disc, where the radiation field is strong and post-shock temperatures exceed $T\ge 10^7~\rm{K}$, inverse Compton cooling by high-energy electrons may briefly compete with collisional losses. As the shock expands and cools, bremsstrahlung dominates during the hot phase \citep{1979rpa..book.....R}. Once the temperature drops into $10^5-10^7~\rm{K}$ range, metal-line cooling becomes dominant \citep{1982ApJ...253..268C,1986ApJ...304..771C}. 

In the present semi-analytic treatment, we approximate the radiative losses using optically thin cooling functions. This is justified as a first-order estimate for the dense post-shock gas, but detailed photoionization and radiation-diffusion effects may modify the exact shell-formation time. During the radiative stage of SNR evolution, the metal-line cooling quickly becomes dominant, leading to the formation of a dense cold shell \citep{2015ApJ...802...99K,2017ApJ...846..133K}. When radiative losses can no longer be neglected, the end time of the ST stage can be estimated as $t_\mathrm{ST}\approx max[1.5t_\mathrm{cool,high},1.5t_\mathrm{cool,low}]$ \citep{2015ApJ...802...99K}, where the cooling time at high temperature is
\begin{equation}
    t_\mathrm{cool,high}=\frac{2.3k_\mathrm{B}T_\mathrm{ST}}{1.2(\gamma+1)n_\mathrm{d}\Lambda_\mathrm{cool,high}(T_\mathrm{ST})},
    \label{eq6}
\end{equation}
where $\Lambda_\mathrm{cool,high}(T_\mathrm{ST})\approx1.4\times10^{-27}T_\mathrm{ST}^{1/2}~\rm{erg~cm^{3}~s^{-1}}$ is the cooling function for $T_\mathrm{ST}>10^7~\mathrm{K}$ \citep{1979rpa..book.....R,2011piim.book.....D}, $n_\mathrm{d}=\rho_\mathrm{d}/(\mu_\mathrm{0} m_\mathrm{p})$ is the number density of the disc, $\gamma=5/3$ is the adiabatic index, and $m_\mathrm{p}$ is the proton mass, and the cooling time at low temperature is
\begin{equation}
    t_\mathrm{cool,low}=\frac{2.3k_\mathrm{B}T_\mathrm{ST}}{1.2(\gamma+1)n_\mathrm{d}Z'\Lambda_\mathrm{cool,low}(T_\mathrm{ST})},
    \label{eq7}
\end{equation}
where $\Lambda_\mathrm{cool,low}(T_\mathrm{ST}) = 1.1\times10^{-22} (T_\mathrm{ST}/10^6{\rm K})^{-0.8}~\rm{erg~cm^{3}~s^{-1}}$ is the cooling function for $10^5~\mathrm{K}<T_\mathrm{ST}<10^7~\mathrm{K}$ \citep{2015ApJ...802...99K,2017ApJ...846..133K}, and $Z'$ is the metallicity in units of the solar value ($Z'\equiv Z/Z_\mathrm{\odot}$; We adopt the solar abundance from \cite{2009ARA&A..47..481A}) and take $Z'=5$ as a fiducial value \citep{2022ApJ...925..121W}.

\textit{PDS stage}: After shell formation, the hot interior bubble still expands due to its thermal pressure. We describe this phase using the modified PDS solution for radiative SNRs \citep{2015ApJ...802...99K}. The average pressure of the hot bubble is
\begin{equation}
    P_\mathrm{Bubble,PDS}=(\gamma-1)\frac{3E_\mathrm{th,PDS}}{4\pi  R_\mathrm{PDS}^{3}},
    \label{eq8}
\end{equation}
where $R_\mathrm{PDS}=R_\mathrm{ST}(t/t_\mathrm{ST})^{2/7}$ is the SNR radius during the PDS stage, and $E_\mathrm{th,PDS}=0.576\times E_\mathrm{SN}({t_\mathrm{ST}}/{t})^{11/7}$ is the thermal energy retained in the hot bubble \citep{2015ApJ...802...99K}. Meanwhile, the temperature of the shell remains roughly the same as the disc temperature at $T_\mathrm{d}$. The swept-up gas is assumed to cool into a geometrically thin shell whose temperature is approximately the local disc temperature $T_\mathrm{d}$. For a thin shell with $\Delta R_\mathrm{shell}\ll R_\mathrm{PDS}$, the shell volume is $V_\mathrm{shell}\simeq4\pi R_\mathrm{PDS}^{2}\Delta R_\mathrm{shell}$. The interior volume is $V_\mathrm{PDS}=4\pi R_\mathrm{PDS}^{3}/3$, so the shell volume fraction is $\chi\equiv{V_\mathrm{shell}}/{V_\mathrm{PDS}}\simeq{3\Delta R_\mathrm{shell}}/{R_\mathrm{PDS}}$. Motivated by numerical simulations, which find typical shell compression factors corresponding to $\chi\sim0.01$-$0.1$ \citep{2021ApJ...906...15M,2024ApJ...965..168R}, we adopt $\chi=0.01$ as the fiducial value. The shell pressure is then estimated as
\begin{equation}
    P_\mathrm{shell}=\frac{k_\mathrm{B}T_\mathrm{d}M_\mathrm{sw}}{V_\mathrm{shell}(\mu_\mathrm{0} m_\mathrm{p})},
    \label{eq9}
\end{equation}
where $M_\mathrm{sw}(R_\mathrm{PDS})=\int_0^{R_\mathrm{PDS}}4\pi R^2\rho_\mathrm{d}dR$ is the swept-up disc mass enclosed by the SNR radius during the PDS stage. The simulations by \cite{2021ApJ...906...15M} indicate that significant shear-driven elongation occurs mainly when the remnant reaches radial widths comparable to several local scale heights. Under our fiducial parameters, the remnant remains smaller than the local disc scale height before the end of the strong expansion phase, suggesting that vertical breakout is not expected to dominate this stage. However, shear-driven deformation and shell instabilities during the subsequent backflow are not resolved in the present model and may affect the efficiency and morphology of seed-cloud formation. The PDS phase ends when the bubble pressure decreases to the shell pressure, $P_\mathrm{Bubble,PDS}=P_\mathrm{shell}$ \citep{2024ApJ...971L..44R}.

Because AGN discs can be optically thick, we also check whether the radiated energy can diffuse out on a timescale shorter than the shell expansion time. We define an effective diffusion time $t_\mathrm{diff}\sim\tau_\mathrm{shell}\Delta R_\mathrm{shell}/c$ and compare it with the local expansion time $t_\mathrm{exp}\sim R_\mathrm{ST}/v_\mathrm{ST}$, where $\tau_\mathrm{shell}\approx\kappa_\mathrm{d}\rho_\mathrm{shell}R_\mathrm{shell}$ is the optical depth of the shell. The results are summarized in Table~\ref{tab3}. Although both the background disc and the compressed shell are optically thick, the fiducial $M_\bullet=10^8M_\odot$ models have $t_\mathrm{diff}/t_\mathrm{exp}<1$ at all three radii. Therefore, radiative energy loss is not trapped on the expansion timescale in these fiducial models, supporting the transition to radiative snowplow evolution. We caution, however, that this conclusion is not universal across the entire parameter grid: in some lower SMBH mass or more optically thick cases, photon diffusion may delay the effective cooling and shift the shell-formation time.

\textit{MCS stage}: After the PDS stage ends, the remnant enters the MCS stage. During this phase, the shell momentum is approximately conserved, and the SNR radius follows $R_\mathrm{MCS}\propto t^{1/4}$with the corresponding shock velocity decreasing as $v_\mathrm{MCS}\propto t^{-3/4}$ \citep{2015ApJ...802...99K,2024ApJ...965..168R}. Assuming that the shell volume fraction remains approximately constant and that the shell temperature scales with the square of the shock velocity, the shell pressure evolves as
\begin{equation}
    P_\mathrm{MCS}=P_\mathrm{shell}(\frac{t}{t_\mathrm{PDS}})^{-3/2},
    \label{eq10}
\end{equation}
where $t_\mathrm{PDS}$ is the end time of the PDS stage. We define the end of the strong SNR expansion as the time when the shell pressure becomes comparable to the ambient disc pressure, $P_\mathrm{MCS}\approx n_\mathrm{d}k_\mathrm{B}T_\mathrm{d}$. This defines $t_\mathrm{MCS}$, when the remnant ceases to drive a strong outward shock, rather than an instantaneous hydrodynamic reversal. In the physical picture adopted here, continued cooling may subsequently make the cavity underpressured relative to the ambient disc, allowing the external pressure gradient to drive an inward-propagating compression wave and backflow \citep{2021ApJ...906...15M,2024ApJ...965..168R}.

\begin{figure}
    \includegraphics[width=0.9\linewidth]{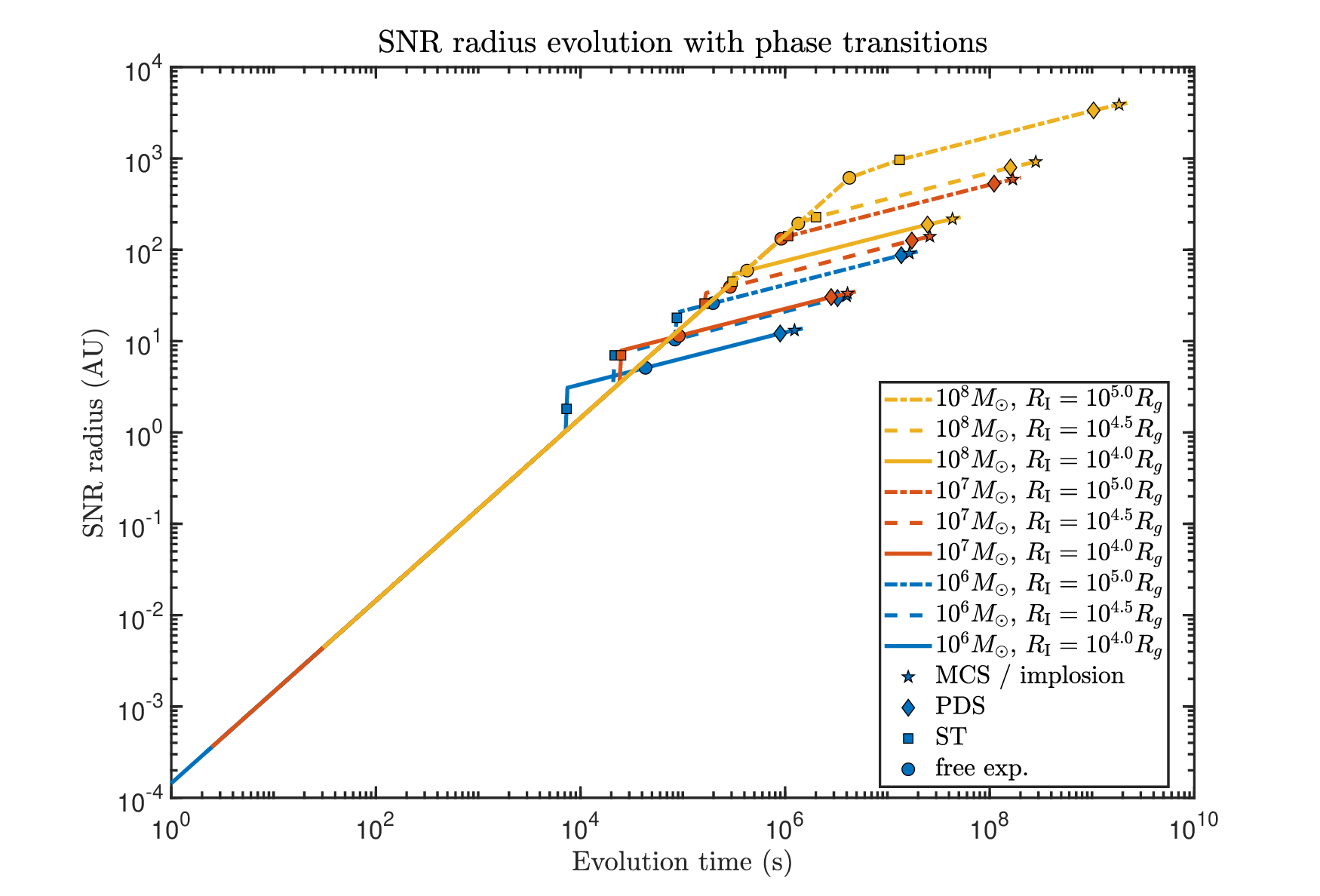}
    \caption{Time evolution of the SNR radius for local spherical models with different SMBH masses and explosion radii. Different colors denote different SMBH masses, while different line styles denote different initial radii $R_\mathrm{I}$. The markers indicate the transition times between the free-expansion, ST, PDS, and MCS stages, and the estimated onset of implosion. For the innermost cases, radiative cooling is efficient, and the ST phase is strongly compressed or effectively skipped. Here, AU denotes Astronomical Units.}
    \label{fig3}
\end{figure}

Figure~\ref{fig3} shows the time evolution of the SNR radius for the local spherical models with different SMBH masses and explosion radii. The curves clearly illustrate the sequence of evolutionary stages, including free expansion, ST stage, PDS stage, MCS stage, and the estimated onset of the subsequent implosion phase. The markers denote the corresponding transition times. Since the radial environmental asymmetry across the remnant is weak, as quantified in Table~\ref{tab1}, we do not show separate inner- and outer-side shock trajectories. Instead, all curves are calculated using the local disc properties evaluated at the explosion radius $R_\mathrm{I}$. At small radii, especially for $R_\mathrm{I}=10^4R_\mathrm{g}$, radiative cooling is very efficient and the cooling time can become shorter than the nominal free-expansion time. In this case, the adiabatic ST stage is strongly compressed or effectively skipped, and the remnant rapidly enters the radiative PDS stage \citep{2019MNRAS.488..978J,2025A&A...695A.271S}. The time $t_\mathrm{MCS}$ marks the end of the strong SNR expansion phase, when the shell pressure becomes comparable to the local ambient gas pressure. It should not be interpreted as a hydrodynamic discontinuity, but rather as the transition point after which the remnant is no longer treated as an expanding SNR. The subsequent shell collapse/backflow and compact seed cloud formation are discussed in the next subsection.

\begin{figure}
    \includegraphics[width=0.9\linewidth]{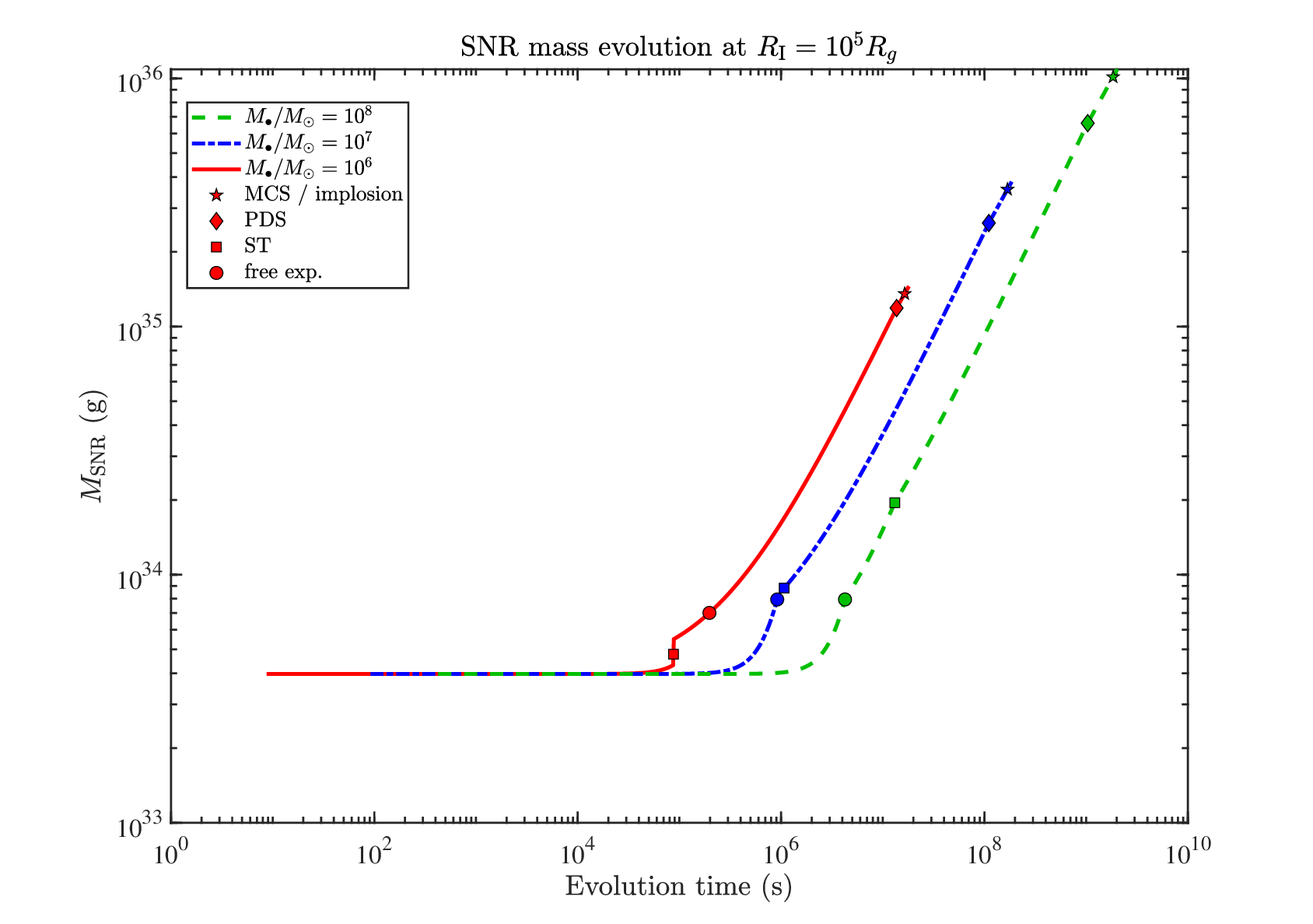}
    \caption{Time evolution of the SNR mass at $R_\mathrm{I}=10^5R_\mathrm{g}$ for different SMBH masses. The initially flat segment corresponds to the ejecta-dominated phase, while the later rapid increase is caused by swept-up disc gas. The markers denote the same evolutionary transition times as in Figure~\ref{fig3}.}
    \label{fig4}
\end{figure}

Figure~\ref{fig4} further shows the SNR mass evolution at $R_\mathrm{I}=10^5R_\mathrm{g}$ for different SMBH masses. The initially flat part corresponds to the ejecta-dominated phase, while the rapid rise at later times reflects the accumulation of swept-up disc gas. The transition markers in the mass evolution are consistent with those in the radius evolution, confirming that the mass growth becomes most significant during the radiative snowplow stages. For the fiducial models considered here, the total SNR evolution time is shorter than the local orbital period, $P_\mathrm{orb}=2\pi/\Omega$, so orbital shear is not expected to dominate the early SNR phase evolution. For a representative stellar density in the self-gravitating outer AGN disc, $n_\star\approx10^4~\rm{pc^{-3}}$ \citep{2024ApJ...967...88C}, the expected number of stars enclosed within the final remnant volume is
\begin{equation}
    N_\star\sim \frac{4\pi}{3}n_\star R_\mathrm{MCS}^{3}\lesssim1,
    \label{eq11}
\end{equation}
indicating that direct stellar encounters during the SNR expansion are unlikely. Vertical stratification of the AGN disc could in principle introduce stronger asymmetry once the shell radius becomes comparable to, or larger than, the local scale height $H_\mathrm{d}$ \citep{2021ApJ...906...15M,2023ApJ...950..161L}. In this work, however, we adopt a midplane local background and focus on the SNR phase evolution before vertical breakout becomes dynamically important. Therefore, chimney-like deformation and vertical breakout are outside the scope of the present model.

Once the outward shock stalls and the cavity becomes underpressured relative to the surrounding disc, an inward-propagating compression wave and associated backflow may develop. Simulations of SNRs in AGN discs show that ambient disc gas can penetrate the weakened remnant and be accelerated inward by the pressure gradient that refills the cavity \citep{2021ApJ...906...15M}. Simulations of late radiative SNR evolution in the ISM additionally show that cold material from the rear side of the shell can flow into the central vacuum and settle into a compact overdensity \citep{2024ApJ...965..168R}. In multidimensional flows, Rayleigh–Taylor and related shell instabilities may fragment the returning layer into clumps or filaments \citep{2000A&A...361..303B}. We therefore use ``implosion" as shorthand for partial and asymmetric backflow, rather than for the homologous contraction of a complete spherical shell. This inward flow is qualitatively analogous to the negative-pressure phase of terrestrial blast waves, although the physical environment and timescales are very different \citep{1977enw..book.....G}.

The backflowing material is expected to move toward the centre of the remnant after the outward SNR expansion ceases. In our semi-analytic model, this process is not followed hydrodynamically. Instead, we estimate the time required for the cooled shell material to return to the remnant centre by introducing a characteristic backflow velocity,
\begin{equation}
    v_\mathrm{back}
    =
    \max
    \left[
    c_\mathrm{s,d},
    |v_\mathrm{MCS}|,
    \left(\frac{2GM_\mathrm{shell}}{R_\mathrm{MCS}}\right)^{1/2}
    \right],
    \label{eq12}
\end{equation}
where $c_\mathrm{s,d}$ is the local sound speed of the disc gas, $v_\mathrm{MCS}$ is the shell velocity at the end of the MCS stage, and the last term represents the characteristic gravitational infall velocity associated with the cold shell mass. Eq.~(\ref{eq12}) provides only an order-of-magnitude, phenomenological estimate of the characteristic backflow speed and does not imply a spatially uniform or spherically coherent inward velocity. Accordingly, $t_\mathrm{backflow}={R_\mathrm{MCS}}/{v_\mathrm{back}}$ should also be regarded as an order-of-magnitude return timescale. Thus, in our model, $t_\mathrm{MCS}$ marks the end of the strong SNR expansion phase, while the compact seed cloud is initialized only after an additional backflow time, $t_\mathrm{implosion}=t_\mathrm{MCS}+t_\mathrm{backflow}$.

During this implosion stage, only a small fraction of the cold shell material is assumed to be transported back to the remnant centre and incorporated into a compact bound core. Recent three-dimensional simulations qualitatively motivate the formation of a central cloud through SNR implosion, but they do not provide a calibrated seed-formation efficiency for AGN-disc conditions \citep{2024ApJ...965..168R,2024ApJ...971L..44R}. We therefore parameterize the initial seed-cloud mass as $M_\mathrm{seed}=f_\mathrm{seed}M_\mathrm{shell}$, and adopt the conservative fiducial value $f_{\rm seed}=10^{-3}$, where the cold shell mass is taken to be $M_\mathrm{shell}=f_\mathrm{shell}M_\mathrm{sw,MCS}$. We adopt $f_\mathrm{shell}=0.8$ as a fiducial shell-mass fraction, motivated by the concentration of swept-up gas into a radiative shell and $M_\mathrm{sw,MCS}$ is the swept-up disc mass at the end of the SNR evolution \citep{1975MNRAS.172...55F}. Here $f_\mathrm{seed}$ is an effective seed-formation efficiency normalized to the shell mass; the seed cloud may contain both cooled shell fragments and entrained or refilling ambient gas.

Since the SNR evolution time is shorter than the local orbital period, the displacement of the remnant centre during the SNR phase is small. We therefore approximate the orbital radius of the newly formed seed cloud as the explosion radius. The compact seed cloud is embedded in the tidal field of the SMBH. Therefore, its maximum equilibrium size is limited by the Hill radius,
\begin{equation}
    R_\mathrm{seed}
    =R_\mathrm{I}
    \left(
    \frac{M_\mathrm{seed}}{3M_\bullet}
    \right)^{1/3}.
    \label{eq13}
\end{equation}
This compact cloud should be regarded as a pressure-confined and tidally limited seed cloud, rather than as a fully self-gravitating cloud at birth. At fixed cloud mass, the Hill radius decreases as the cloud is transported inward. If inward advection or orbital migration occurs faster than self-gravitating collapse, the seed cloud may be tidally stripped, and its material may eventually be incorporated into the inner accretion flow. The coupled evolution of cloud migration, tidal disruption, and collapse is not followed in the present model. Its subsequent growth, survival, and collapse are then followed by considering accretion from the surrounding disc gas and by checking the relevant gravitational, tidal, shear, ionization, and magnetic criteria. 

The resulting backflow time is not negligible compared with the late SNR evolution time. For the fiducial $M_\bullet=10^8M_\odot$ models, we find $t_\mathrm{backflow}\simeq5.5$-$233~\mathrm{yr}$ and $t_\mathrm{implosion}=t_\mathrm{MCS}+t_\mathrm{backflow}\simeq6.9$-$292~\mathrm{yr}$. Thus, the seed cloud is not assumed to appear instantaneously at $t_\mathrm{MCS}$; instead, $t_\mathrm{MCS}$ marks the termination of the expanding SNR phase, while the compact seed cloud is initialized only after one characteristic backflow time.

\subsection{Cloud evolution and SF}
\label{subsec: The evolution of the cloud and star formation}

\begin{table}
\centering
\small
\renewcommand{\arraystretch}{1.25}
\begin{tabular}{lcccc}
\hline
$R_\mathrm{I}$ ($R_\mathrm{g}$)
& $\tau_\mathrm{d}$
& $\tau_\mathrm{shell}$
& $t_\mathrm{diff}/t_\mathrm{exp}$
& Cooling \\
\hline
$10^{4}$ 
& $\sim6.6\times10^{3}$ 
& $\sim7.7\times10^{2}$ 
& 0.365 
& efficient \\
$10^{4.5}$ 
& $\sim9.5\times10^{2}$ 
& $\sim82.8$ 
& 0.0253 
& efficient \\
$10^{5}$ 
& $\sim1.4\times10^{2}$ 
& $\sim9.4$ 
& 0.00185 
& efficient \\
\hline
\end{tabular}
\caption{Cooling diagnostics for the fiducial $M_\bullet=10^8M_\odot$ models. The disc is optically thick, but the diffusion time remains shorter than the expansion time, supporting efficient radiative energy loss during shell formation. Auxiliary calculations indicate that cooling becomes less robust in part of the lower-SMBH-mass parameter space.}
\label{tab3}
\end{table}

The seed cloud is initially pressure confined and tidally limited, and subsequently evolves toward true self-gravitating collapse through continued accretion. The formation of the SNR bubble breaks the local axisymmetry of the disc. Therefore, the subsequent gas supply to the seed cloud is better described by local Hill capture in a sheared flow than by the global viscous inflow of the background disc. Consequently, the mass growth of the cloud cannot be governed by the global, axisymmetric disc viscosity \citep{1973A&A....24..337S}.

Instead, the accretion is physically limited by the cloud's gravitational reach within the strongly sheared disc environment, characterized by its Hill radius $R_\mathrm{H}(t) = R_\mathrm{I}(t)(M_\mathrm{cl}(t)/3M_\bullet)^{1/3}$. For a dense cloud entirely embedded within the disc ($R_\mathrm{H} \lesssim H_\mathrm{d}$), the accretion is effectively 3D. The relative velocity of the disc gas entering the Hill sphere is dominated by the Keplerian shear, $\Delta v \approx \Omega R_\mathrm{H}$. The mass accretion rate into the Hill sphere can be explicitly derived as:
\begin{equation}
\dot{M}_\mathrm{acc} \approx \pi f_\mathrm{cav}\rho_\mathrm{d} (\Omega R_\mathrm{H}) R_\mathrm{H}^2 = \pi f_\mathrm{cav}\rho_\mathrm{d} \Omega R_\mathrm{H}^3,
\label{eq14}
\end{equation}
where $f_\mathrm{cav}$ accounts for the reduced gas density inside the SNR cavity, and $f_\mathrm{cav}=1$ serves as the fiducial value for the model; it should be interpreted as the efficient-refilling limit rather than a generic post-SNR cavity state. By substituting the definition of the Hill radius, the $R_\mathrm{H}^3$ term explicitly cancels the cube root of the cloud mass, yielding:
\begin{equation}
\dot{M}_\mathrm{acc} \approx \pi f_\mathrm{cav}\rho_\mathrm{d} \Omega R_\mathrm{I}^3 \left(\frac{M_\mathrm{cl}}{3M_\bullet}\right) = \frac{M_\mathrm{cl}}{t_\mathrm{growth}},
\label{eq15}
\end{equation}
where $t_\mathrm{growth} = 3M_\bullet / (\pi f_\mathrm{cav}\rho_\mathrm{d} \Omega R_\mathrm{I}^3)$ is the shear-limited growth timescale. This derivation shows that the cloud mass undergoes exponential growth before gravitational collapse. However, the mechanism is driven by local 3D Hill capture in a shear flow, rather than global viscous angular momentum transport. Thus, before collapse, the cloud mass grows approximately exponentially,
\begin{equation}
    M_\mathrm{cl}(t)
    =
    M_\mathrm{seed}
    \exp
    \left[
    \frac{t-t_\mathrm{implosion}}{t_\mathrm{growth}}
    \right],
    \label{eq16}
\end{equation}
where $t_\mathrm{implosion}=t_\mathrm{MCS}+t_\mathrm{backflow}$ is the time at which the seed cloud is initialized.

We parameterize the cloud radius as
\begin{equation}
    R_\mathrm{cl}
    =
    R_\mathrm{seed}
    \left(
    \frac{M_\mathrm{cl}}{M_\mathrm{seed}}
    \right)^a,
    \label{eq17}
\end{equation}
where $R_\mathrm{seed}$ is the initial cloud radius defined in Eq.~(\ref{eq13}), and the index $a$ controls how rapidly the cloud expands while accreting mass. The limiting case $a=0$ corresponds to an approximately fixed-radius cloud \citep{1975ApJ...195..715M,1977ApJ...211..135C}, while $a=1/3$ corresponds to a nearly constant-density cloud \citep{2003MNRAS.341..501S,2007MNRAS.374..515L}. Choosing $a=1/8$ (lying within the physically motivated regime $0<a<1/3$) implies that the cloud radius increases only sub-linearly with mass, while the mean density continues to grow as $\rho_\mathrm{cl}\propto M_\mathrm{cl}^{5/8}$. This behavior physically describes a cold gas clump that contracts under the combined action of its increasing self-gravity and the immense ambient gas pressure of the AGN disc. This mechanism is mathematically analogous to the standard core-growth models in planetary gas accretion theory \citep{1996Icar..124...62P}, where the gas envelope contracts within the Hill sphere while steadily drawing material from the surroundings.

\begin{figure}
    \centering
    \includegraphics[width=0.9\linewidth]{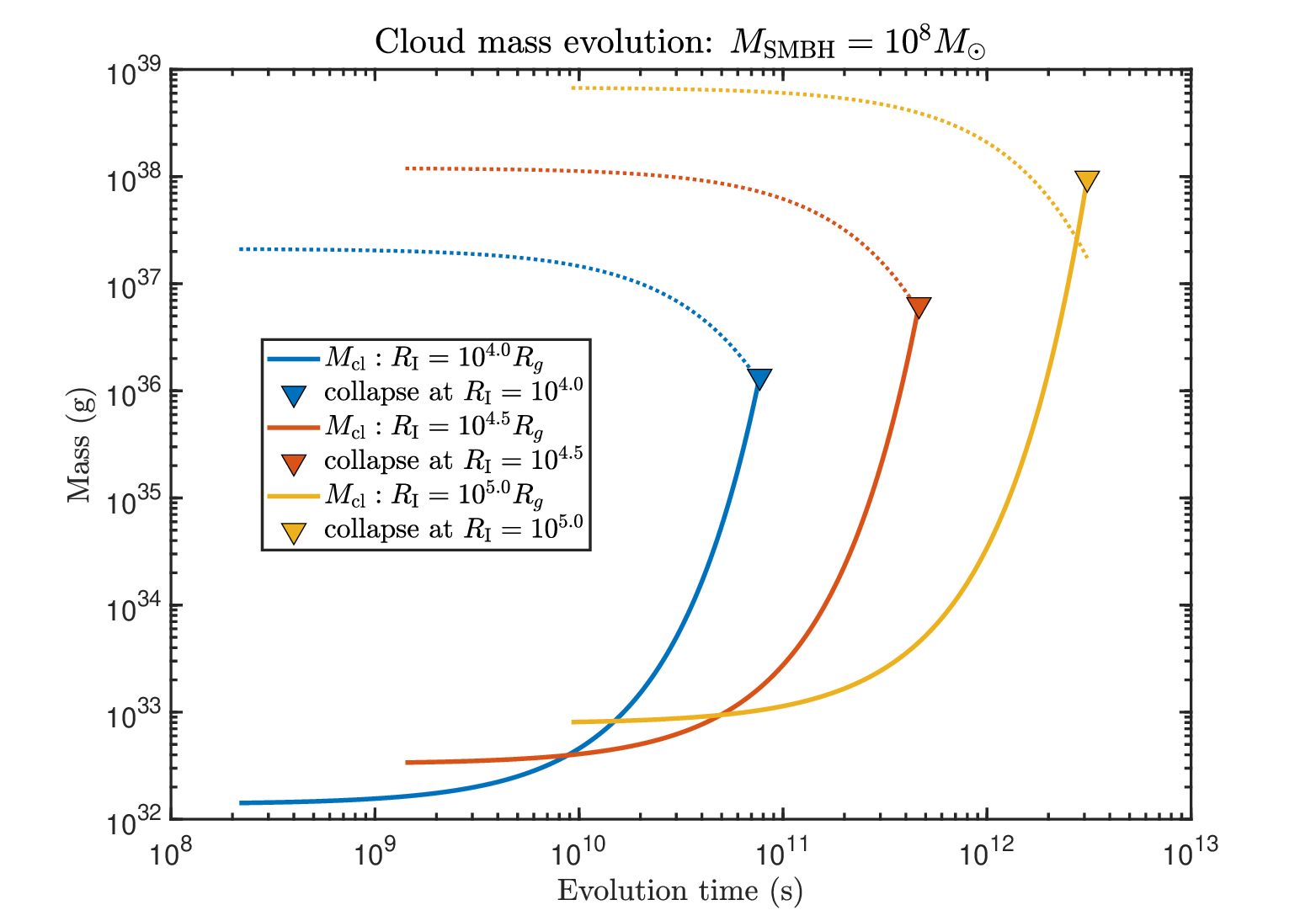}
    \caption{Cloud mass evolution for the fiducial SMBH mass $M_\bullet=10^8M_\odot$ at three explosion radii. Solid curves show the cloud mass $M_\mathrm{cl}$, while dotted curves show the corresponding Jeans mass $M_\mathrm{Jeans}$. Downward triangles mark the first time at which all collapse criteria are simultaneously satisfied. The collapse time and collapse mass increase with $R_\mathrm{I}$ because the ambient disc density decreases and the Hill-capture growth time becomes longer at larger radii. These values correspond to the fiducial cloud-growth model with $f_\mathrm{cav}=1$, and $a=1/8$.}
    \label{fig5}
\end{figure}

\begin{table*}
\small
\renewcommand{\arraystretch}{1.30}
\centering
\begin{tabular}{llcccccccc}
\hline
$M_\bullet$ ($M_\odot$)
& $R_\mathrm{I}$ ($R_\mathrm{g}$)
& $M_\mathrm{cl}$ ($M_\odot$)
& $M_{\star,\mathrm{event}}$ ($M_\odot$)
& $M_{\star,\mathrm{event}}/M_\mathrm{shell}$
& $N_{\mathrm{total}}$
& $N_{0.1-1}$
& $N_{1-8}$
& $N_{8-100}$
& $f_{>8M_\odot}$ \\
\hline

\multirow{3}{*}{$10^{6}$}
& $10^{4}$
& 2.34
& 0.702
& 0.047
& 0.049
& 0.016
& 0.015
& 0.018
& 36.6\% \\

& $10^{4.5}$
& 3.23
& 0.969
& 0.037
& 0.067
& 0.022
& 0.020
& 0.024
& 36.6\% \\

& $10^{5}$
& 7.43
& 2.23
& 0.042
& 0.154
& 0.051
& 0.046
& 0.056
& 36.6\% \\

\hline

\multirow{3}{*}{$10^{7}$}
& $10^{4}$
& 16.08
& 4.82
& 0.188
& 0.334
& 0.111
& 0.100
& 0.122
& 36.6\% \\

& $10^{4.5}$
& 62.96
& 18.89
& 0.344
& 1.31
& 0.435
& 0.393
& 0.478
& 36.6\% \\

& $10^{5}$
& $3.24\times10^{2}$
& 97.07
& 0.684
& 6.71
& 2.24
& 2.02
& 2.45
& 36.6\% \\

\hline

\multirow{3}{*}{$10^{8}$}
& $10^{4}$
& $6.87\times10^{2}$
& $2.06\times10^{2}$
& 2.885
& 14.24
& 4.75
& 4.29
& 5.21
& 36.6\% \\

& $10^{4.5}$
& $2.85\times10^{3}$
& $8.54\times10^{2}$
& 5.640
& 59.08
& 19.69
& 17.78
& 21.60
& 36.6\% \\

& $10^{5}$
& $4.85\times10^{4}$
& $1.46\times10^{4}$
& 35.84
& $1.01\times10^{3}$
& $3.36\times10^{2}$
& $3.03\times10^{2}$
& $3.68\times10^{2}$
& 36.6\% \\

\hline
\end{tabular}
\caption{Expected numbers of stars in different mass ranges per SNR-induced cloud-collapse event for different SMBH masses and explosion radii. Here $M_\mathrm{cl}$ is the cloud mass at the first collapse point, and $M_{\star,\mathrm{event}}=f_\mathrm{\star}M_\mathrm{cl}$ with $f_\mathrm{\star}=0.3$. The stellar numbers should be strictly interpreted as statistical expectation values per event rather than deterministic counts for an individual event, especially for the low-mass configurations where $M_\mathrm{cl}$ is small. For this fixed IMF slope, the number fraction of stars with $M>8M_\odot$ is independent of $M_\bullet$ and $R_\mathrm{I}$, giving $f_{>8M_\odot}=36.6\%$. These values correspond to the fiducial cloud-growth model with $f_\mathrm{cav}=1$, $a=1/8$, and $f_\star=0.3$.}
\label{tab4}
\end{table*}

The cloud density is $\rho_\mathrm{cl}={3M_\mathrm{cl}}/{4\pi R_\mathrm{cl}^{3}}$. We first diagnose self-gravity using the Jeans mass and the virial parameter. The Jeans mass is
\begin{equation}
    M_\mathrm{Jeans}
    =
    \left(
    \frac{\pi c_\mathrm{s,d}^{2}}{G}
    \right)^{3/2}
    \rho_\mathrm{cl}^{-1/2},
    \label{eq18}
\end{equation}
and the virial parameter is
\begin{equation}
    \alpha_\mathrm{vir}
    =
    \frac{5c_\mathrm{s,d}^{2}R_\mathrm{cl}}
    {GM_\mathrm{cl}} .
    \label{eq19}
\end{equation}
Here $c_\mathrm{s,d}$ is the local sound speed of the ambient disc gas evaluated at $R_\mathrm{I}$. As $M_\mathrm{cl}$ grows, the cloud density increases, causing $M_\mathrm{Jeans}$ and $\alpha_\mathrm{vir}$ to decrease. However, self-gravity alone is not sufficient for collapse in the AGN disc environment. We also require the cloud to remain confined against the SMBH tidal field, survive local shear, avoid photoionization-driven disruption, and become magnetically supercritical. We identify the true gravitational collapse point only when the growing cloud simultaneously overcomes all stabilizing and disruptive environmental factors, satisfied by the joint criteria: $M_\mathrm{cl}\ge M_\mathrm{Jeans}, \alpha_\mathrm{vir}\le2, R_\mathrm{cl}\le R_\mathrm{H}, t_\mathrm{ff}\le t_\mathrm{shear}, U_\mathrm{AGN}\le U_\mathrm{crit}, \lambda_\mathrm{B}\ge1$. Here $t_\mathrm{ff}=\left({3\pi}/{32G\rho_\mathrm{cl}}\right)^{1/2}$
is the free-fall time, $t_\mathrm{shear}=2/(3\Omega)$ is the local shear time. The photoionization criterion is expressed in terms of the AGN ionization parameter \citep{1999agnf.book.....K,2006agna.book.....O},
\begin{equation}
    U_\mathrm{AGN}
    =
    \frac{Q_\mathrm{AGN}}
    {4\pi R_\mathrm{I}^{2}n_\mathrm{cl}c},
    \label{eq20}
\end{equation}
where $Q_\mathrm{AGN}\sim 10^{54} ~\rm{photons~s^{-1}}$ is the ionizing photon production rate of the AGN, consistent with standard spectral energy distribution (SED) templates \citep{1994ApJS...95....1E}. We only use the fixed baseline ionization photon rate as a diagnostic indicator for local survival conditions. For small-mass SMBHs, this value may be a conservative estimate. $n_\mathrm{cl}={\rho_\mathrm{cl}}/{\mu_0m_\mathrm{p}}$ is the cloud number density. We adopt $U_\mathrm{crit}=10^{-2}$ as a fiducial diagnostic threshold for photoionization-driven disruption \citep{1999agnf.book.....K,2006agna.book.....O}. This value should be regarded as an order-of-magnitude survival criterion rather than a sharply calibrated boundary. Thus, the condition $U_\mathrm{AGN}\le U_\mathrm{crit}$ ensures that the cloud is sufficiently dense to resist AGN photoionization \citep{1990ApJ...354..529B}.

The magnetic condition is described by the dimensionless mass-to-flux ratio,
\begin{equation}
    \lambda_\mathrm{B}
    =
    \frac{2\pi G^{1/2}M_\mathrm{cl}}
    {\Phi_\mathrm{B}},
    \label{eq21}
\end{equation}
where the magnetic flux threading the cloud is approximated as $\Phi_\mathrm{B}\simeq\pi R_\mathrm{cl}^{2}B_\mathrm{cl}.$ As an order-of-magnitude estimate, we characterize the local magnetic field by $B_\mathrm{d}=(2\dot{M}_\mathrm{\bullet}c/R_\mathrm{I}^2 )^{1/2}$ \citep{1999ASPC..190..173L,2017NewAR..79....1L}, where $\dot{M}_{\bullet}$ is the accretion rate of the SMBH, and assume flux freezing during cloud compression, giving
\begin{equation}
    B_\mathrm{cl}
    =
    B_\mathrm{d}
    \left(
    \frac{\rho_\mathrm{cl}}{\rho_\mathrm{d}}
    \right)^{2/3},
    \label{eq22}
\end{equation}
which should account only for magnetic-field amplification by ideal flux-freezing compression. Shock-driven turbulence may additionally amplify the field through a small-scale dynamo, thereby decreasing $\lambda_\mathrm{B}$ and delaying or suppressing collapse in marginal cases \citep{2015MNRAS.450.4035F}. Conversely, turbulent reconnection and reconnection diffusion may remove magnetic flux from the accumulating cloud, weaken magnetic support, and increase $\lambda_\mathrm{B}$ \citep{2010ApJ...714..442S}. Compared with the normal ISM, the AGN-disc models considered here have much higher gas density and pressure and are subject to strong differential rotation, which may sustain magnetohydrodynamic (MHD)  turbulence and magnetic-field amplification or regeneration \citep{1998RvMP...70....1B,2003MNRAS.341..501S}. We regard the present magnetic criterion as a baseline diagnostic rather than a complete MHD prediction. Therefore,
\begin{equation}
    \lambda_\mathrm{B}
    =
    \frac{2\pi G^{1/2}M_\mathrm{cl}}
    {\pi R_\mathrm{cl}^{2}B_\mathrm{cl}} .
    \label{eq23}
\end{equation}
The condition $\lambda_\mathrm{B}\ge1$ means that the cloud is magnetically supercritical and can collapse against magnetic support \citep{1976ApJ...210..326M,2012ARA&A..50...29C}.

In the evolutionary tracks, the calculation is stopped at the first point satisfying all of the above conditions, and the corresponding time and mass are recorded as $t_\mathrm{collapse}$ and $M_\mathrm{collapse}$. We do not explicitly follow the subsequent SF feedback or post-collapse mass-loss phase.

\begin{figure*}
    \centering
    \includegraphics[width=0.9\linewidth]{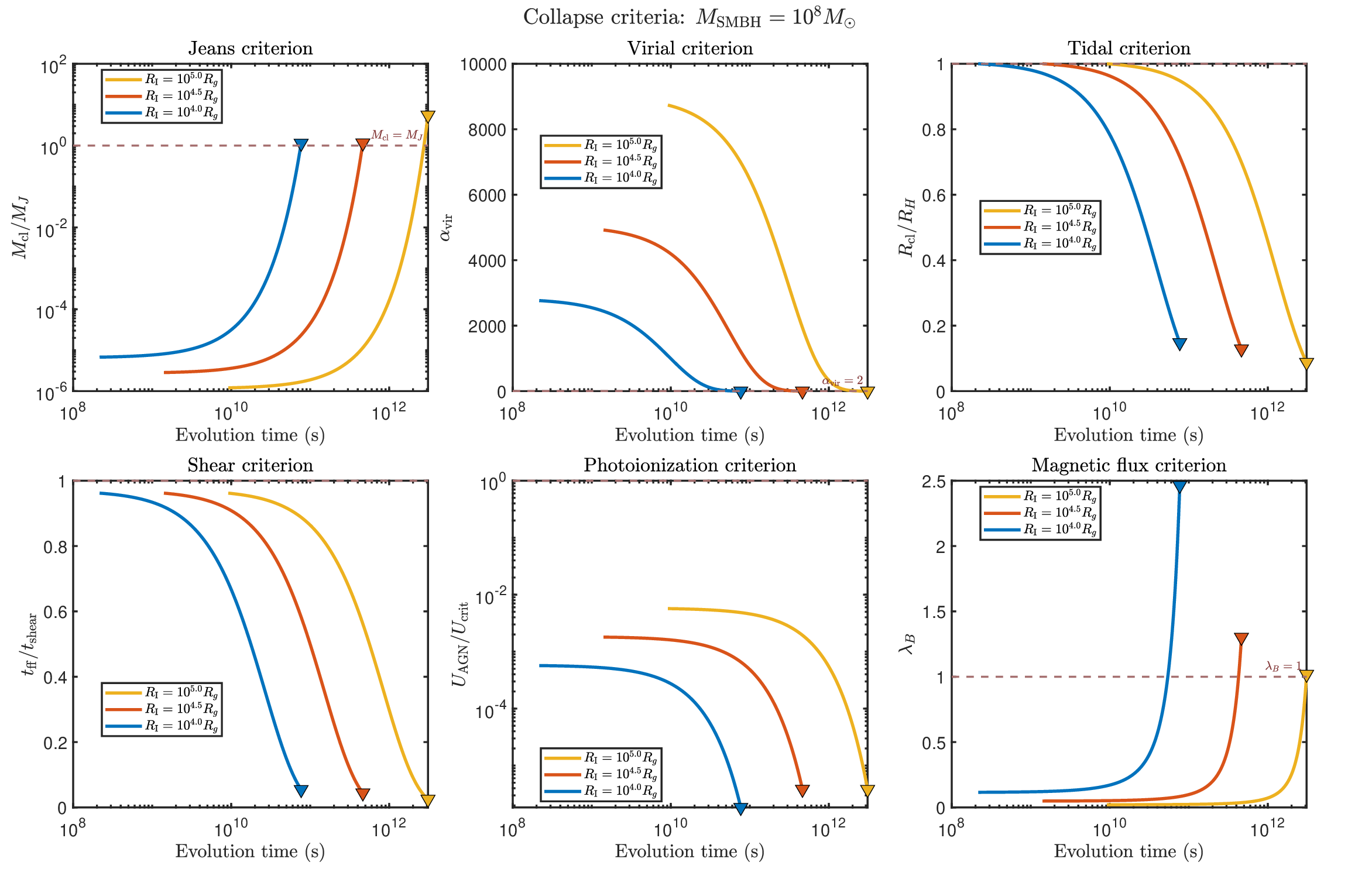}
    \caption{Time evolution of the individual collapse and survival criteria for seed clouds embedded in the AGN disc, shown for the fiducial SMBH mass $M_\bullet=10^8M_\odot$ and three explosion radii. The panels show the Jeans ratio $M_\mathrm{cl}/M_\mathrm{Jeans}$, virial parameter $\alpha_\mathrm{vir}$, tidal ratio $R_\mathrm{cl}/R_\mathrm{H}$, shear ratio $t_\mathrm{ff}/t_\mathrm{shear}$, photoionization parameter $U_\mathrm{AGN}/U_\mathrm{crit}$, and magnetic mass-to-flux ratio $\lambda_\mathrm{B}$. The horizontal dashed lines mark the corresponding critical values, and the downward triangles indicate the collapse points identified in Figure~\ref{fig5}.}
    \label{fig6}
\end{figure*}

Figure~\ref{fig5} shows the cloud mass evolution for the fiducial SMBH mass $M_\bullet=10^8M_\odot$ at three explosion radii. The solid curves show the exponential growth of $M_\mathrm{cl}$ after seed cloud initialization, while the dotted curves show the corresponding Jeans mass. The downward triangles mark the first time at which the full set of collapse criteria is satisfied. As $R_\mathrm{I}$ increases, the ambient disc density decreases and the Hill-capture growth time becomes longer. Consequently, both the collapse time and the collapse mass increase with distance from the SMBH.

Figure~\ref{fig6} further decomposes the collapse condition into the individual criteria. The Jeans ratio $M_\mathrm{cl}/M_\mathrm{Jeans}$ rises toward unity, while the virial parameter decreases toward $\alpha_\mathrm{vir}\le2$, confirming that the growing cloud progressively becomes self-gravitating. At the same time, the ratios $R_\mathrm{cl}/R_\mathrm{H}$ and $t_\mathrm{ff}/t_\mathrm{shear}$ remain below unity at collapse, indicating that the cloud is tidally confined and collapses faster than it is disrupted by local shear. The photoionization parameter also remains well below the critical value for the cases considered here, so AGN photoionization does not prevent collapse. The magnetic criterion, represented by $\lambda_\mathrm{B}$, increases during accretion and reaches the supercritical regime near the collapse point. Thus, the final collapse is not determined by a single Jeans crossing alone, but by the simultaneous satisfaction of the gravitational, tidal, shear, photoionization, and magnetic conditions.

Considering the IMF as $\xi(M)=AM^{-\Gamma}$, where $\Gamma\simeq1$ is adopted as the fiducial index of a top-heavy IMF in AGN discs \citep{2022MNRAS.512.2573T}, the normalization constant is determined by the stellar mass formed from the collapsed cloud,
\begin{equation}
    A=
    \frac{f_\mathrm{\star}(M_\mathrm{cl}/M_\odot)}
    {\int_{0.1}^{100}M^{1-\Gamma}\,dM}.
    \label{eq24}
\end{equation}
Here the total stellar mass formed in one cloud-collapse event is $M_{\star,\mathrm{event}}=f_\mathrm{\star}M_\mathrm{cl}$, and $f_\mathrm{\star}=0.3$ is the assumed integrated cloud-to-star conversion efficiency within the dense collapsed core, which is consistent with the empirical core-to-star efficiency in nearby dense star-forming regions, as inferred from the offset between the prestellar core mass function and the stellar IMF \citep{2007A&A...462L..17A,2014prpl.conf...27A}. The mass $M_\mathrm{cl}$ is taken from the cloud mass at the first collapse point in our fiducial cloud-evolution calculation.

The expected number of stars in a mass interval is then
\begin{equation}
    N_{M_1-M_2}=\int_{M_1}^{M_2}\xi(M)\,dM.
    \label{eq25}
\end{equation}
For the fiducial top-heavy IMF with $\Gamma=1$, the number fraction of stars more massive than $8M_\odot$ is
\begin{equation}
    f_{>8M_\odot}
    =
    \frac{\ln(100/8)}{\ln(100/0.1)}
    \simeq 36.6\%.
    \label{eq26}
\end{equation}
If the IMF slope varies within $0.5\leq\Gamma\leq1.0$ as suggested by \cite{2022MNRAS.512.2573T}, this number fraction ranges from $\simeq74.1\%$ to $\simeq36.6\%$. Therefore, the absolute number of second-generation SNe is sensitive to the adopted IMF slope. However, as long as the IMF remains top-heavy, the stellar mass and feedback budget are still dominated by massive stars. For example, the mass fraction in stars above $8M_\odot$ is $\simeq92\%$ for $\Gamma=1$ and $\simeq98\%$ for $\Gamma=0.5$. Our conclusions, therefore, rely mainly on the dominance of massive stars in the top-heavy IMF, while the exact number of massive stars should be regarded as IMF-dependent.

We emphasize that $f_\star$ is an assumed integrated cloud-to-star conversion factor used to normalize the IMF. It is not a prediction of the present model. The quantity directly predicted by our cloud evolution is the collapse mass $M_\mathrm{collapse}$. We therefore characterize the event-level stellar yield as
\begin{equation}
    Y_{\star,\mathrm{event}}
    \equiv
    \frac{M_{\star,\mathrm{event}}}{M_\mathrm{shell}}
    =
    f_\star
    \frac{M_\mathrm{collapse}}{M_\mathrm{shell}} .
    \label{eq27}
\end{equation}
Because the cloud can continue to grow by Hill accretion after seed-cloud formation, $Y_{\star,\mathrm{event}}$ can exceed unity. It should therefore be interpreted as the stellar mass produced per SNR-triggered cloud-collapse event relative to the initial shell mass, rather than as an instantaneous star-formation efficiency.

Using the collapse masses obtained for each SMBH mass and explosion radius, we estimate the expected stellar population formed in each SNR-induced cloud-collapse event. The results are summarized in Table~\ref{tab4}. Since the IMF slope is fixed to the fiducial top-heavy value $\Gamma=1$, the relative number fractions in different stellar mass ranges are the same for all models. The variation across the table, therefore, mainly reflects the dependence of the collapse mass on $M_\bullet$ and $R_\mathrm{I}$. For the same fiducial $M_\bullet=10^8M_\odot$ models, the stellar yield per initial shell mass is $M_{\star,\mathrm{event}}/M_\mathrm{shell}\simeq2.9$-$35.8$, reflecting that the cloud continues to grow by Hill accretion after the initial seed-cloud formation.

\subsection{Cloud-collapse sensitivity to $f_\mathrm{cav}$ and $a$}
\label{sec:sen}

In the previous section, we defined the fiducial seed-cloud evolution and the full set of collapse criteria, including the Jeans, virial, tidal, shear, photoionization, and magnetic conditions. Here we do not repeat these criteria. Instead, we examine how the collapse outcome depends on the two least constrained parameters in the post-SNR cloud-growth model: the cavity density factor $f_\mathrm{cav}$ and the cloud mass-radius index $a$.

The parameter $f_\mathrm{cav}$ accounts for the reduced gas density inside the SNR cavity after the shell implosion. In the shear-limited Hill-capture prescription, the cloud-growth timescale is characterized by $t_\mathrm{growth}$ (defined in Eq.~(\ref{eq15})). Therefore, a smaller $f_\mathrm{cav}$ directly lengthens the cloud-growth time and delays the moment at which the cloud can satisfy the collapse criteria. Physically, this corresponds to a more strongly evacuated cavity, in which less gas is available for accretion onto the seed cloud.

\begin{table}
\centering
\small
\renewcommand{\arraystretch}{1.25}
\begin{tabular}{lccc}
\hline
$f_\mathrm{cav}$ 
& Collapse fraction
& Physical interpretation \\
\hline
$1$ 
& 1.00
& Efficient cavity refilling \\
$10^{-2}$ 
& 0.944
& Moderately depleted cavity \\
$10^{-4}$ 
& 0.472
& Strongly depleted cavity \\
\hline
\end{tabular}
\caption{Summary of the collapse fraction in the cloud-sensitivity calculation. The result shows that the fiducial collapse channel is robust when the post-SNR cavity is not strongly depleted, but becomes sensitive to the gas-refilling efficiency for small $f_\mathrm{cav}$.}
\label{tab5}
\end{table}

The index $a$ controls how the cloud radius changes during accretion (i.e., Eq.~(\ref{eq17})). The corresponding cloud density scales as $\rho_\mathrm{cl}\propto M_\mathrm{cl}^{1-3a}$. Thus, for $a<1/3$, the cloud becomes denser as it accretes mass. Smaller values of $a$ lead to faster density growth, a more rapid decrease in the Jeans mass and virial parameter, and hence earlier collapse. In contrast, larger values of $a$ allow the cloud radius to expand more efficiently during accretion, delaying collapse and sometimes preventing the cloud from satisfying all collapse criteria within the adopted integration time.

\begin{figure*}
    \centering
    \includegraphics[width=0.9\linewidth]{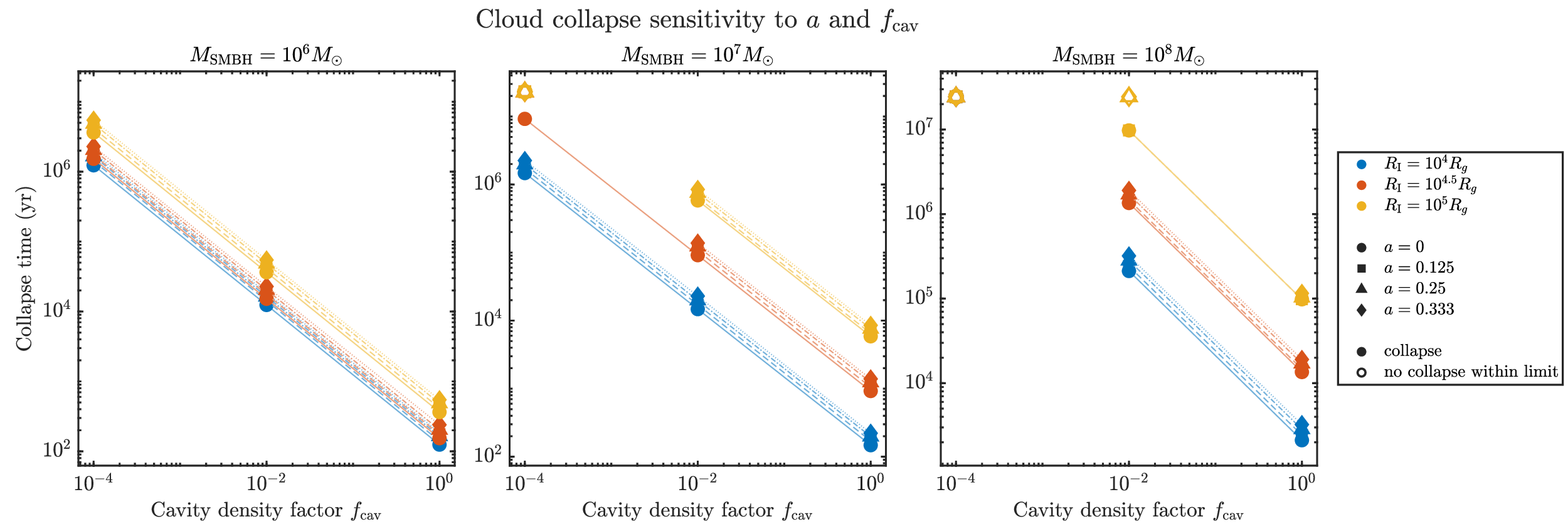}
    \caption{Sensitivity of the seed-cloud collapse time to the cavity density factor $f_\mathrm{cav}$ and the cloud mass-radius index $a$. The three panels correspond to $M_\bullet=10^6$, $10^7$, and $10^8~M_\odot$. Colors indicate the initial radius $R_\mathrm{I}=10^4$, $10^{4.5}$, and $10^5~R_\mathrm{g}$, while different marker shapes indicate $a=0$, $1/8$, $1/4$, and $1/3$. Filled symbols denote models that satisfy all collapse criteria, whereas open symbols indicate models that do not collapse within the adopted integration time and are plotted near the upper edge for visualization only.}
    \label{fig7}
\end{figure*}

Figure \ref{fig7} shows the resulting collapse time as a function of $f_\mathrm{cav}$ for three SMBH masses and three explosion radii. The dominant trend is that the collapse time decreases strongly with increasing $f_\mathrm{cav}$. This behaviour follows directly from $t_\mathrm{growth}\propto f_\mathrm{cav}^{-1}$: when the cavity is strongly depleted, the cloud accretes more slowly and requires a much longer time to become self-gravitating. For the most depleted cases, especially at larger radii and higher SMBH masses, some models do not collapse within the adopted integration time.

At fixed $f_\mathrm{cav}$ and $M_\bullet$, clouds initialized at larger $R_\mathrm{I}$ generally collapse later. This is because the ambient disc density decreases outward, reducing the Hill-capture supply rate and increasing the growth time. The effect of the mass-radius index $a$ is secondary but systematic: larger $a$ usually produces later collapse, while smaller $a$ leads to earlier collapse because the cloud density increases more rapidly during mass growth. The separation between different $a$ values is most visible when $f_\mathrm{cav}$ is small, where the collapse outcome is already close to the no-collapse boundary.

Overall, Figure \ref{fig7} shows that the fiducial collapse channel is robust when the cavity is not strongly depleted, but it becomes sensitive to the gas-refilling efficiency when $f_\mathrm{cav}\ll1$. The most favorable conditions for prompt collapse are therefore small $R_\mathrm{I}$, large $f_\mathrm{cav}$, and small-to-moderate $a$. Conversely, large radii, strongly evacuated cavities, and large $a$ values delay collapse or suppress it within the integration window.

Table~\ref{tab5} shows that the collapse outcome is controlled primarily by the amount of gas available in the post-SNR cavity. When $f_\mathrm{cav}=1$, all models in the explored grid collapse. For $f_\mathrm{cav}=10^{-2}$, the collapse fraction remains high, but for $f_\mathrm{cav}=10^{-4}$ it drops to about one half. Thus, SNR-triggered seed-cloud collapse is not automatic; it requires sufficient gas refilling or retention inside the post-SNR cavity.

The event-level stellar yield can be compared with other AGN-disc star-formation channels in an order-of-magnitude way. If the embedded CCSNe rate is $\Gamma_\mathrm{CCSNe}$, the stellar mass production rate associated with the present channel is
\begin{equation}
    \dot{M}_{\star,\mathrm{SNR}}
    \sim
    \Gamma_\mathrm{CCSNe}
    M_{\star,\mathrm{event}} .
    \label{eq28}
\end{equation}
As shown in Table \ref{tab4}, for the fiducial $M_\bullet=10^8M_\odot$ models, $M_{\star,\mathrm{event}}$ ranges from $\sim2\times10^2M_\odot$ to $\sim1.5\times10^4M_\odot$ per SNR-induced cloud-collapse event. Therefore, this mechanism is unlikely to replace global gravitational fragmentation of the AGN disc as the primary star-formation channel in all systems. Instead, it should be regarded as a localized recycling channel, in which a CCSN can convert part of the swept-up and refilled cavity gas into a new compact cloud and, under favourable conditions, trigger a second generation of massive stars. As a consistency check, the local gas reservoir in the surrounding disc is sufficiently large. For the fiducial outer-disc case, the gas mass contained in a narrow annulus of width $\Delta R\sim0.01R_\mathrm{I}$ is already of order $\gtrsim 10^5M_\odot$, exceeding the largest collapse mass in Table \ref{tab4}. Therefore, the large final cloud masses should not be interpreted as originating from the initial shell alone, but from subsequent Hill capture of the local disc reservoir.

\section{Summary \& Discussion}
\label{sec: Summary}

In this work, we investigate whether SN remnants (SNRs) embedded in AGN discs can trigger the formation of compact seed clouds and subsequently lead to second-generation SF. We adopt the SG disc model and consider SMBH masses of $M_\bullet=10^6$, $10^7$, and $10^8M_\odot$, with representative explosion radii $R_\mathrm{I}=10^4$, $10^{4.5}$, and $10^5R_\mathrm{g}$. The AGN disc properties are evaluated locally at the explosion radius, and the subsequent SNR and cloud evolution are followed using these local disc conditions.

We first examine the radial environmental asymmetry across the expanding remnant. For the fiducial case shown in Table \ref{tab1}, and for the grid checked in our calculation, the density, pressure, and temperature contrasts are modest. Therefore, the radial pressure gradient of the AGN disc does not qualitatively change the stage-by-stage SNR evolution. In the main calculation, we consequently model the remnant as a local spherical SNR expanding into a uniform medium whose properties are evaluated at $R_\mathrm{I}$.

For the fiducial $M_\bullet=10^8M_\odot$ models, the cooling and diffusion diagnostics indicate that radiative losses are efficient on the shell-expansion timescale. In the inner disc, the nominal ST stage can be strongly compressed or effectively skipped, and the remnant rapidly enters the radiative pressure-driven and MCS stages. The total SNR evolution time is shorter than the local orbital period in the fiducial models, so orbital shear does not dominate the early SNR expansion. At the end of the strong expansion phase, the shell pressure becomes comparable to the local ambient gas pressure. We then treat the subsequent shell collapse/backflow as the channel that initializes a compact, pressure-confined and tidally limited seed cloud.

The seed cloud grows by shear-limited Hill capture from the surrounding AGN disc gas. In this prescription, the accretion rate is regulated by the gas density available in the SNR cavity, parameterized by $f_\mathrm{cav}$, and by the cloud mass-radius relation, $R_\mathrm{cl}\propto M_\mathrm{cl}^{a}$. We identify cloud collapse only when the Jeans, virial, tidal, shear, photoionization, and magnetic mass-to-flux criteria are simultaneously satisfied. For the fiducial model with $a=1/8$ and $f_\mathrm{cav}=1$, the seed cloud can survive the AGN disc environment, grow through Hill accretion, and eventually become self-gravitating. The collapse time and collapse mass generally increase with explosion radius because the ambient disc density decreases outward.

Using the cloud mass at the first collapse point, we estimate the expected stellar population for a top-heavy IMF with $\Gamma=1$ and an integrated cloud-to-star conversion efficiency $f_\star=0.3$. The expected number of massive stars is strongly dependent on $M_\bullet$ and $R_\mathrm{I}$. In low-SMBH-mass cases, the expected number of massive stars per event can be below unity, while in the $M_\bullet=10^8M_\odot$ models the same mechanism can produce from several to hundreds of massive stars, depending on the explosion radius. Thus, SNR-induced cloud collapse in AGN discs provides a plausible pathway for localized second-generation SF, especially in massive SMBH systems and in regions where the post-SNR cavity is not strongly depleted.

We have not explicitly followed the subsequent dynamics of compact remnants formed in core-collapse events. In general, CCSNe may leave behind neutron stars or stellar-mass black holes, and anisotropic explosions can impart natal kicks to these compact remnants \citep{1994Natur.369..127L,1996PhRvL..76..352B,2013MNRAS.434.1355J,2017ApJ...837...84J,2018ApJ...865...61G}. Observations and population studies suggest that neutron stars can receive kicks of hundreds of $\mathrm{km~s^{-1}}$, while black hole kicks may range from very small values to tens or hundreds of $\mathrm{km~s^{-1}}$, depending on the explosion asymmetry and fallback mass \citep{2012MNRAS.425.2799R,2013MNRAS.434.1355J,2024PhRvL.132s1403V}. Such kicked compact remnants could perturb the local SNR shell or the later seed-cloud environment, especially after the remnant has entered the radiative snowplow stage and the shell has become dense and slow. In addition, accretion onto a newly formed sBH may provide local radiative or mechanical feedback, which could either heat, compress, or partially disrupt nearby cold gas. These processes involve multidimensional hydrodynamics, compact-object motion, fallback/accretion physics, and radiation feedback, and are therefore beyond the scope of the present semi-analytic model. In this work, we focus on the SNR-driven formation and growth of seed clouds using the local AGN disc environment, and regard compact-remnant kicks and accretion feedback as additional complications to be explored in future numerical simulations.

We must acknowledge several important physical simplifications inherent in our semi-analytical approach. First, regarding the photoionization criterion ($U_\mathrm{AGN}$), a rigorous treatment would require full radiation-transfer calculations considering local shielding and the dynamic photoevaporative mass loss of the cloud surface, which could compete with the Hill accretion rate \citep{1990ApJ...354..529B}. Second, the magnetic criterion ($\lambda_\mathrm{B}$) relies on ideal flux-freezing. Finally, our results reveal a steep dependence on the SMBH mass (Table 4), with the mechanism becoming largely inefficient in low-mass ($\sim 10^6 M_\mathrm{\odot}$) systems. This implies that our channel cannot serve as a universal mechanism to explain the global metal enrichment across all active galaxies. Instead, it represents a localized gas recycling channel. We isolate the dynamic feasibility of Hill-capture growth in this work, and leave the complex interplay of radiation-hydrodynamics, dynamic photoevaporation, and dynamic MHD dynamos to future high-resolution three-dimensional radiation-MHD simulations.

Several physical effects remain to be explored in future work. We do not explicitly follow the 3D deformation of the SNR caused by vertical disc stratification, turbulence, or strong shear, nor do we model detailed radiation-MHD cooling and magnetic-field amplification during shell compression. Multiple SNe occurring in the same star-forming region could also change the shell mass, cavity structure, and subsequent cloud growth. In addition, if the second-generation stellar population produces new massive stars and eventually new sBHs, their feedback, dynamics, and natal kicks may further complicate the local environment. These effects could either disrupt the cloud or trigger additional episodes of SF. Fully assessing this later nonlinear evolution will require high-resolution 3D simulations. Nevertheless, within the local model and parameter space considered here, our results show that CCSNe embedded in AGN discs can produce compact seed clouds that survive, grow, and collapse, making second-generation SF a possible outcome.

\section*{Acknowledgments}

We thank the anonymous referee for constructive suggestions and comments, and Ken Chen, Bao-Quan Huang, and Xiao-Yan Li for helpful discussions. This work was supported by the National Key R\&D Program of China under grant 2023YFA1607902, the National Natural Science Foundation of China under grants 12494572 and 12221003, and the Fund of National Key Laboratory of Plasma Physics (No. 6142A04240201).

\section*{Data availability}

The data underlying this article will be shared on reasonable request to the first author.

\label{lastpage}
\end{document}